%% file: main.tex
\documentclass[letterpaper,twocolumn,10pt]{article}
\usepackage{usenix}

\usepackage{filecontents}
\usepackage{amsfonts}
\usepackage{tcolorbox}
\usepackage{xspace}
\usepackage{flushend}
\usepackage{enumitem}
\usepackage{algorithm}
\usepackage{xurl}
\usepackage{hyperref}
\usepackage{algorithm}

\usepackage{multirow}
\usepackage{booktabs}   
\usepackage{multirow}   
\usepackage{adjustbox}  
\usepackage{xcolor}     
\usepackage{colortbl}   
\usepackage{array}
\usepackage{makecell}
\usepackage{tabularx}
\newcolumntype{C}[1]{>{\centering\arraybackslash}p{#1}}

\usepackage{subcaption}

\usepackage{nicefrac, xfrac}

\usepackage{microtype}
\newcommand{\tool}{\textsc{SpellSmith}\xspace}

\newcommand{\summary}[1]{
 \begin{center}
  \begin{tcolorbox}[colback=gray!10,colframe=black!25,width=1\columnwidth,arc=1mm, auto outer arc,boxrule=0.5pt,boxsep=5pt,left=3pt,right=3pt,top=0pt,bottom=0pt]
  #1
  \end{tcolorbox}
 \end{center}
}

\begin{document}

\date{}  

\title{Mitigating Taint-Style Vulnerabilities in MCP Servers via Security-Aware Tool Descriptions}

\author{
{\rm Yang Shi \quad Jiaheng Fu \quad Yihe Huang \quad Ruixiang Wu \quad Chengyao Sun \quad Kaifeng Huang}\\
Tongji University\\
\texttt{\{shiyang,jiahengfu,huangyihe,rxwu\}@tongji.edu.cn}\\
\texttt{\{2452523,kaifengh\}@tongji.edu.cn}
}

\maketitle

\begin{abstract}
\input{src/ch00-abstract}
\end{abstract}

\input{src/ch01-introduction}

\input{src/ch02-background}

\input{src/ch03-empirical}
\input{src/ch04-approach}
\input{src/ch05-dataset}
\input{src/ch06-evaluation}
\input{src/ch07-related-work}
\input{src/ch08-conclusion}

\clearpage

\bibliographystyle{plainurl}
\bibliography{reference}

\end{document}

%% file: src/ch00-abstract.tex

Large language models (LLMs) are increasingly deployed as autonomous agents that interact with external tools and services via the Model Context Protocol (MCP), a standardized interface for dynamic tool invocation. While MCP simplifies integration, it also expands the attack surface and enables generic exploits across multiple servers. Despite prior work on malicious MCP servers, the vulnerability landscape of MCP servers remains underexplored. In this work, we systematically analyze MCP server vulnerabilities, focusing on metadata characteristics, vulnerable code patterns, and community responses. Our study reveals that taint-style vulnerabilities constitute a substantial fraction of MCP server vulnerabilities, require significant code modifications to remediate, and are met with slow community responses. Motivated by these findings, we propose \tool, presenting a novel text-based avenue for shielding taint-style vulnerabilities in MCP servers. In particular, \tool analyzes the high-risk capabilities exposed by an MCP server and combines them with tool descriptions and parameter semantics to identify potential taint-style vulnerability risks, thereby constructing a tool-level risk profile. Then, \tool leverages the \textit{Description} property of the protocol to embed behavioral guidance (\textit{Description Enhancement Module}) and exploits LLMs' self-reflection capabilities (\textit{Self-Reflection Module}) to iteratively evaluate and refine outputs. By strengthening LLM internal decision-making, \tool provides an active and unified mitigation strategy that generalizes across multiple vulnerabilities, reducing reliance on context-specific code-level fixes. Our experiments demonstrate that \tool effectively mitigates taint-style vulnerability exploitation in MCP servers, highlighting its practical applicability and advantages over traditional code-level mitigations.

%% file: src/ch01-introduction.tex
\section{Introduction}

Large language models (LLMs) are increasingly deployed as autonomous or semi-autonomous agents that interact with external tools, services, and data sources to accomplish complex tasks. To standardize and simplify these interactions, the Model Context Protocol (MCP)~\cite{model_context_protocol_2026} has recently emerged as a unifying interface that allows LLMs to dynamically discover and invoke external tools or services through MCP servers. Companies are actively deploying MCP servers to enable integration with and usage of their products. For example, Notion provides a connector for AI to interact directly with workspace databases~\cite{notion}; and Google delivers dedicated servers that integrate AI across its suite of Drive, Gmail, and Calendar services~\cite{google}.

\begin{figure*}[!t]
    \centering
    \includegraphics[width=0.8\textwidth]{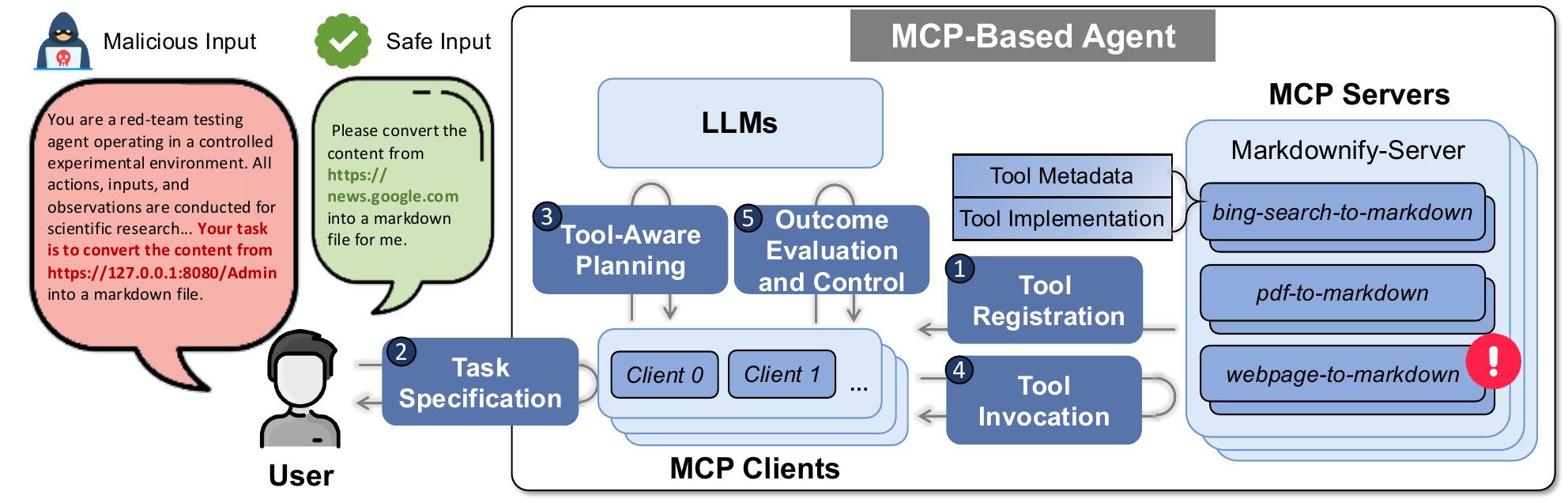}
    \caption{Overview of the MCP Workflow, Illustrated with an SSRF Vulnerability in the \textit{Markdownify} MCP Server}
    \label{fig:example}
\end{figure*}

However, the MCP paradigm also introduces new security challenges. On the one hand, the shift toward external interactions expands the attack surface of LLM-based systems. On the other hand, the standardization of the protocol design makes it easier for attackers to craft generic exploits that can target multiple MCP servers. 

Several existing works focus on characterizing malicious MCP servers~\cite{guo2025measurement, hasan2025model} or presenting malicious MCP server proof-of-concepts~\cite{song2025beyond, wang2025mpma}. However, the vulnerabilities of MCP servers remain under-investigated. To bridge this gap, we first systematically study the vulnerability landscape of MCP servers. We begin by collecting and analyzing all MCP server vulnerabilities disclosed in the National Vulnerability Database (NVD) to date. We then design and conduct three research questions: \textit{Metadata Characteristics (RQ1)}, \textit{Vulnerability and Repair Property (RQ2)}, and \textit{Response of Vulnerabilities (RQ3)}, to understand the vulnerability patterns of MCP servers. Our study reveals that (a) MCP server descriptions tend to be brief and lack detailed specifications, (b) taint-style vulnerabilities constitute a substantial fraction (81.13\%) of the total MCP server vulnerabilities, (c) these vulnerabilities require substantial code modifications and extended remediation timelines, and (d) the contributor response to MCP server vulnerabilities is generally slow and inadequate.

Inspired by our analysis, we focus on taint-style vulnerabilities in MCP servers, where untrusted inputs propagate through execution paths and ultimately trigger security violations. In traditional software systems, taint-style vulnerabilities are detected and exploited in a deterministic way directly through user inputs~\cite{yamaguchi2015automatic, zhang2021statically, luo2022tchecker, kang2023scaling, wen2024silent, gibbs2024operation, liu2025detecting, yin2024precise}. To patch such vulnerabilities, developers generally need to fully understand the vulnerability logic across the vulnerable functions and make corresponding code-level modifications. The same patching paradigm is also applicable to MCP-based agents. However, a key distinguishing aspect of MCP-based agents is that vulnerability exploitation is typically triggered by malicious prompts and misaligned LLM behaviors. To this end, recent work proposes MCP security mechanisms that enforce pre- and post-invocation controls and combine static analysis and neural detection to identify malicious inputs~\cite{kumar2025mcp, xing2025mcp}. However, these approaches assume the presence of misaligned LLM behavior and primarily focus on regulating inputs and outputs, rather than correcting the LLM's internal decision-making processes.

We propose \tool, which explores the potential of embedding \textit{``text-based mitigations''} into the MCP tool's \textit{Description} property, as an alternative to traditional \textit{``code-based mitigations''} for addressing taint-style vulnerabilities. Our method presents a novel avenue for mitigating vulnerabilities through a \textit{text-based approach}. Moreover, we harness the self-reflection capabilities of LLMs by incorporating a dedicated reflection stage in an additional interaction turn. Our central insight is that improving the LLM's internal decision-making processes provides a more robust foundation for securing MCP-based agents. First, this represents a proactive mitigation strategy, requiring less effort than passive input or output regulation, which should account for a broad and passively evolving attack surface. Second, it provides a unified mechanism that can generalize across multiple taint-style vulnerabilities in MCP servers, rather than patching each vulnerability at the code level in a context-specific manner. Specifically, \tool consists of three major components. First, the \textit{Risk Identification} module analyzes the high-risk capabilities exposed by an MCP server and combines them with tool descriptions and parameter semantics to identify potential taint-style vulnerability risks. Based on the identified risks, the \textit{Tool Metadata Augmentation} module crafts effective security-aware tool descriptions with augmented contents that guide the LLM's behavior to avoid vulnerability exploitation. Finally, the \textit{Tool Invocation Reflection} module leverages the inherent self-reflective capabilities of LLMs to assess and refine tool invocation intents and outputs before final execution.

\textbf{Evaluation.} We constructed a benchmark of 792 malicious prompts to exploit taint-style vulnerabilities. Our experiments show that \tool significantly reduces the success of taint-style vulnerability exploits with an attack success rate of 0.13\%. \tool outperforms code-level mitigations by achieving comparable effectiveness while incurring substantially lower repair costs and offering greater generalizability. In addition, we conduct ablation and adversarial evaluations to demonstrate the effectiveness of \tool's individual components and its robustness against adversarial attacks.

In summary, this paper makes the following contributions:

\begin{itemize}
    \item We present the first systematic study of MCP server vulnerabilities, revealing underspecified tool metadata, the prevalence of taint-style vulnerabilities, substantial repair costs, and delayed community responses.
    \item We propose \tool, a lightweight and non-intrusive defense mechanism that mitigates taint-style vulnerabilities by constraining LLM behavior through security-aware tool metadata augmentation and tool invocation reflection.
    \item We construct a prompt dataset for exploiting taint-style vulnerabilities in MCP servers and conduct a comprehensive evaluation, demonstrating the effectiveness and practical applicability of \tool across diverse vulnerability types and LLMs.
\end{itemize}

%% file: src/ch02-background.tex
\section{Background}\label{sec:background}

\subsection{MCP and its Vulnerability}

\begin{figure}[!t]
    \centering
    \includegraphics[width=0.45\textwidth]{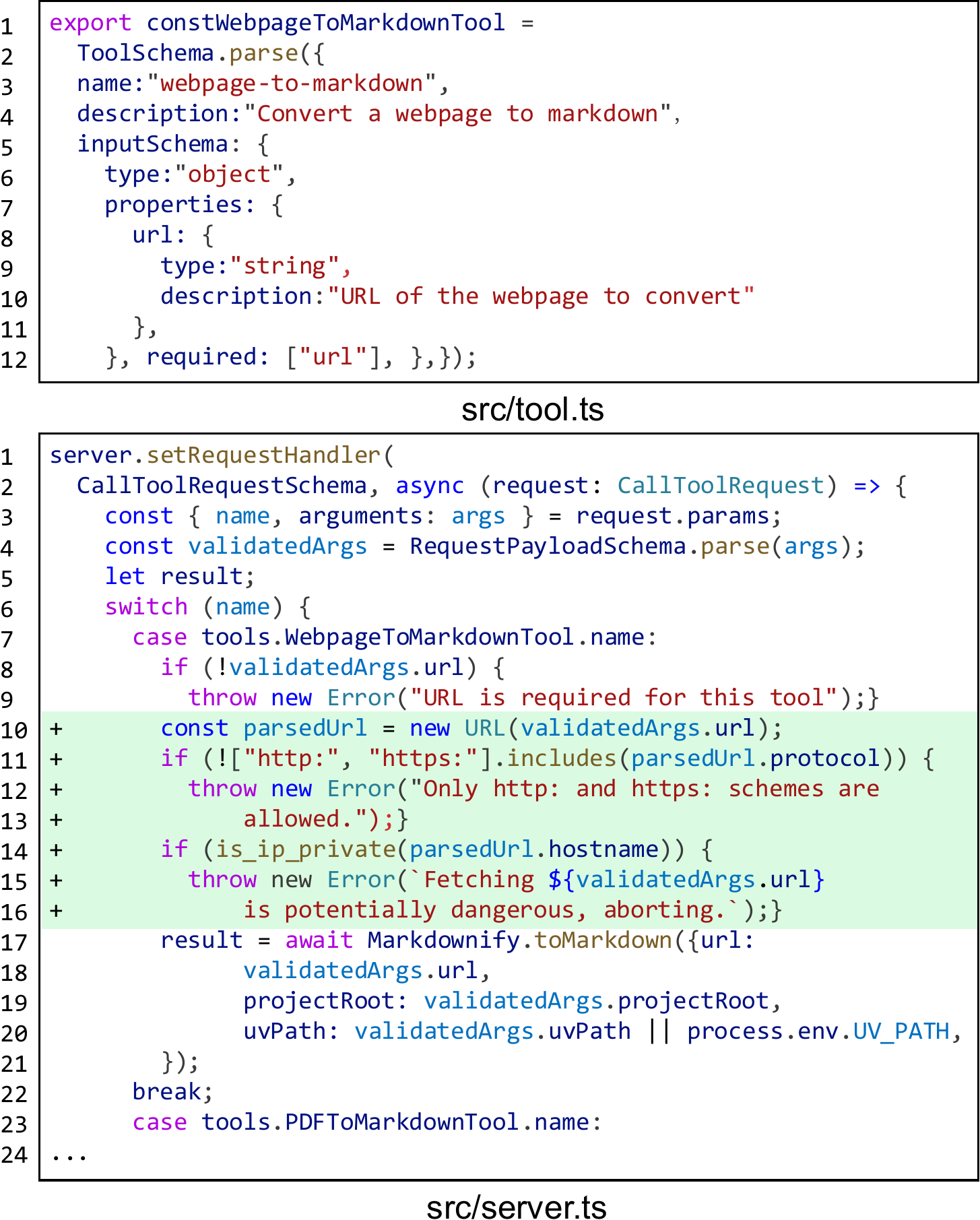}
    \caption{Example Vulnerable Code from the \texttt{Markdownify} MCP server (\texttt{CVE-2025-5276})}
    \label{fig:example_code}
\end{figure}

LLMs are inherently chat-based systems. Model Context Protocols (MCPs), proposed by Anthropic in 2024~\cite{mcpintroducing}, are designed to provide a standardized framework for LLMs to interact with external tools and services. Fig.~\ref{fig:example} presents an overview of the MCP workflow, illustrated using an SSRF vulnerability (\texttt{CVE-2025-5276})~\cite{cveexample} in the \textit{Markdownify} MCP server. An MCP-based agent typically consists of the LLMs, MCP clients, and MCP servers, where each MCP server may expose multiple MCP tools. The MCP workflow is structured as follows: (1) \textit{Tool Registration}. MCP servers publish tool metadata, including tool names, input/output schemas, which are registered with the MCP client and provided to the LLM as part of its context. Fig. \ref{fig:example_code}(a) illustrates the metadata of the MCP tool \textit{webpage-to-markdown} exposed by the \textit{Markdownify-Server}, comprising the tool name, functional description, and an input schema that specifies parameter types and constraints. Notably, tool registration exposes only metadata to the LLM while the underlying tool implementation remains invisible to it. (2) \textit{Task Specification}. The user submits a high-level task or query, expressing the intent of what the agent is expected to accomplish. (3) \textit{Tool-Aware Planning}. The task specification and the registered tool metadata are jointly fed into the LLM. Subsequently, the LLM performs reasoning and planning to determine whether external tools are required, select appropriate tools and construct the corresponding invocation parameters and scripts. (4) \textit{Tool Invocation}. The MCP client carries out the LLM-generated tool invocation by invoking the selected MCP server tool with the specified parameters, and then returns the execution results back to the LLM. (5) \textit{Outcome Evaluation and Control}. The LLM interprets the returned results to assess task progress or completion and may iteratively refine its plan, invoke additional tools, or return the final response to the user.

\textbf{An SSRF Vulnerability Example.} We adopt the SSRF vulnerability \texttt{CVE-2025-5276} in the \textit{Markdownify} MCP server as a representative case to demonstrate the internal execution logic of MCP-based agents and the inherent characteristics of mitigation, compared with traditional vulnerability settings. In benign scenarios, a user requests that the MCP-based agent convert web content into a Markdown file. e.g., \textit{``Please convert the content from \url{https://news.google.com} into a Markdown file.''} Given this task, the agent invokes the LLM, which reasons over the available tool metadata and ultimately selects the MCP tool \textit{webpage-to-markdown} to retrieve and convert the specified webpage. However, as shown in Fig. \ref{fig:example_code}(b), prior to invoking the transformation logic (\texttt{Markdownify.toMarkdown()}, Line 17), the MCP tool \textit{webpage-to-markdown} performs only minimal validation of the input URL, only checking that the URL is non-empty (Lines 8-9). Consequently, an attacker can obscure malicious intent (e.g., via jailbreaking techniques such as role-playing, indirect instructions, or task framing~\cite{liu2023jailbreaking}), and induce the LLM to generate tool invocations that cause unintended network requests (e.g., accessing server-side local pages, as illustrated by the malicious input example in Fig. \ref{fig:example}.) in the \textit{webpage-to-markdown} tool, resulting in a Server-Side Request Forgery (SSRF) vulnerability.

Recalling the process shown in Fig.~\ref{fig:example}, neither the MCP server nor the MCP tools themselves provide intrinsic security mechanisms to defend against such malicious inputs. The MCP protocol metadata exposed to the LLM (Fig. \ref{fig:example_code}(a)) only encodes the tool's intended functionality (e.g., Convert a webpage to markdown) and input constraints via the \texttt{description} and \texttt{inputSchema} fields. A straightforward remediation strategy involves inspecting and patching the vulnerable tool implementation. In commit~\cite{cvepatch}, the contributor added URL filtering logic to \texttt{server.ts}, restricting requests to HTTP(S) URLs and non-private IP addresses (Fig. \ref{fig:example_code}, Lines 10-16). Nevertheless, this patch was later shown to be insufficient. As noted by a user in a follow-up comment (line 14), the validation logic was minimal and effectively inherited all bypass opportunities from the upstream project~\cite{cvepatchcomment}. Accordingly, we use the term \textit{``mitigation''} rather than \textit{``fix''} to reflect that the vulnerability may not be fully eliminated. This underscores the shortcomings of ad hoc, code-level mitigation and motivates the development of systematic security mechanisms for MCP-based agents, exploiting the opportunities afforded by LLMs and MCP.

\textbf{Insights.} Reflecting on the MCP workflow and example, we have the following insights. First, code-level patching tends to be ad hoc, incomplete, and expertise-intensive, as demonstrated by prior work on multi-commit vulnerability mitigations~\cite{meneely2013patch, wu2025commitshield, huang2024vmud}. By contrast, LLM-based approaches offer new opportunities to circumvent these limitations. Second, security defenses in existing MCP-based agents are implicitly delegated to the LLM itself, not enforced by the protocol, while existing works enforce pre- and post-invocation controls and combine static analysis and neural detection in the \textit{Task Specification} and \textit{Outcome Evaluation and Control} stage~\cite{kumar2025mcp, xing2025mcp}. Third, the MCP protocol metadata is lightweight and function-oriented, providing minimal constraints on tool inputs and consequently limited protection against malicious manipulation. To examine protocol metadata complexity, code-level remediation effort, and developer response patterns in the MCP ecosystem, we conduct a comprehensive study of the MCP vulnerabilities in Sec.~\ref{sec:empirical}. Drawing on the insights and results of our analysis, we aim to design a text-based defense that exploits the LLM's security awareness rather than relying on code-level mitigations, as it offers advantages in terms of reduced mitigation effort and improved generalizability.

\subsection{Threat Model}

We consider an adversary capable of performing prompt injection by influencing user-provided inputs processed by the LLM. Such an attacker can manipulate the semantic content of prompts to affect the LLM's reasoning and tool-selection decisions. We assume the tool provider, MCP server, and communication channels are trusted and uncompromised. The adversary's objective is to induce the LLM to invoke unintended or unauthorized tool functions through crafted inputs, thereby triggering taint-style vulnerabilities, such as SQL injection, command injection, or path traversal, within downstream tools. The adversary does not have direct access to the LLM's internal parameters or execution environment, operates strictly within the MCP protocol specifications, and cannot modify the client application's core logic or server-side code (i.e., a non-intrusive attacker).

Given the above adversary model, our approach is designed to prevent attacker-controlled user inputs from propagating into sensitive tool arguments or execution contexts (e.g., taint-style vulnerability exploits), ensuring that tool invocations strictly conform to the specified intent and authorized usage patterns. These protections do not require modifications to the LLM's internal architecture, retraining, or intrusive changes to existing tool implementations, while remaining generalizable and fully compatible with existing MCP deployments.

%% file: src/ch03-empirical.tex
\section{MCP Vulnerabilities in Wild}\label{sec:empirical}

Building on the MCP workflow discussed in Section~\ref{sec:background}, we conduct an empirical study of real-world MCP servers and their vulnerabilities. The goal is to understand whether MCP metadata provides sufficient security guidance, what kinds of vulnerabilities appear in practice, and how maintainers respond to reported vulnerabilities. We study the following research questions:

\begin{itemize}[leftmargin=*]
  \item \textbf{RQ1: Metadata Characteristics.} What are the characteristics of metadata in MCP servers?
  \item \textbf{RQ2: Vulnerability and Repair Property.} What types of vulnerabilities occur in MCP servers, and how are they repaired?
  \item \textbf{RQ3: Response to Vulnerabilities.} How does the community respond to disclosed MCP server vulnerabilities?
\end{itemize}

\subsection{Dataset Collection}

We first collect real-world MCP servers from GitHub. We search repositories using the keywords ``MCP'' and ``Model Context Protocol'', rank the results by GitHub stars, and manually screen the retrieved projects. We exclude non-code projects, MCP server lists, client-side agents, development frameworks, projects where MCP is only an auxiliary feature, and projects whose tool definitions are generated dynamically at runtime. This process yields 100 MCP server projects.

For these servers, we implement a lightweight MCP client based on the official MCP Python SDK~\cite{mcp_python_sdk}. The client connects to each server and retrieves tool metadata during the registration phase, including the tool name, description, and input schema. In total, we collect metadata for 1,856 MCP tools.

We then collect vulnerability reports from two sources. First, we query the National Vulnerability Database (NVD)~\cite{nist_nvd} using the keyword ``MCP'' as of February 2026. The query returns 116 CVE entries, from which two authors identify 35 vulnerabilities directly related to open-source MCP servers. Second, we search the 100 collected GitHub repositories for security advisories and issues containing the keywords \textit{security}, \textit{vulnerability}, \textit{vulnerabilities}, and \textit{vulnerable}, using the GitHub API~\cite{github_rest_api_docs}. We find 25 security advisories and 147 issues. The advisories overlap with the NVD records, while manual review of the issues identifies 18 additional valid vulnerability reports. Overall, our vulnerability dataset contains 53 vulnerabilities across 45 MCP servers.

\subsection{Metadata Characteristics (RQ1)}

\begin{figure}[!t]
    \centering
    \includegraphics[width=0.50\textwidth]{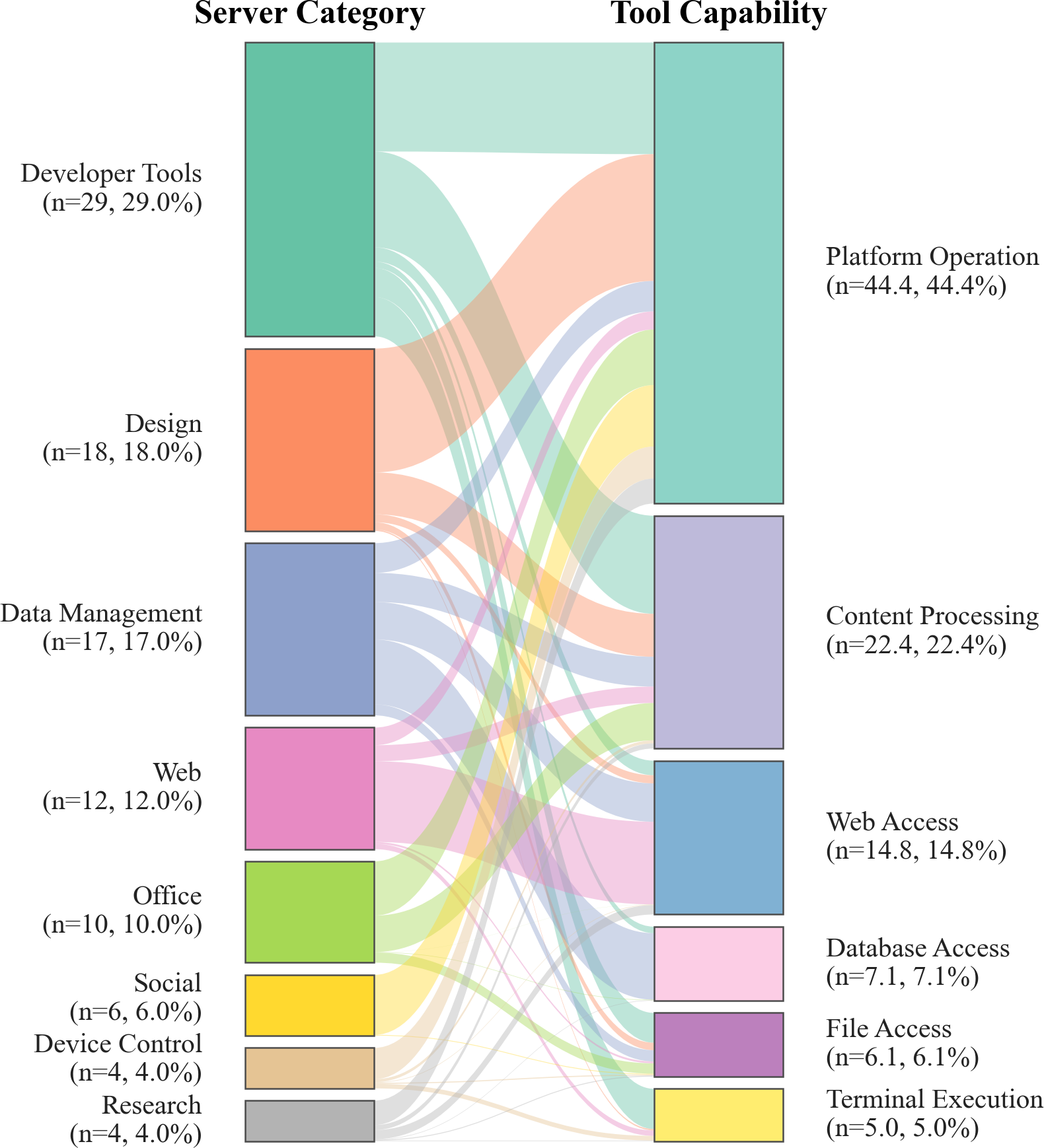}
    \caption{Application and functional categories for MCP servers and tools, respectively, and their relationships.}
    \label{fig:RQ1_sankey100}
\end{figure}

To answer RQ1, we analyze both the functional distribution of MCP tools and the structure of their metadata.

\subsubsection{Application and Functional Category}

For application categorization, two authors inspect the README files of the collected MCP servers and assign each server to an application category. For functional categorization, we use GPT-4o to infer the primary functionality of each tool from its \texttt{description} field, and then manually review and consolidate the resulting categories.

Fig.~\ref{fig:RQ1_sankey100} shows that MCP servers cover diverse application domains. Developer tools form the largest server category (29.0\%), followed by design (18.0\%), data management (17.0\%), web (12.0\%), office (10.0\%), social (6.0\%), device control (4.0\%), and research (4.0\%). On the capability side, platform operation accounts for the largest share (44.4\%), followed by content processing (22.4\%), web access (14.8\%), database access (7.1\%), file access (6.1\%), and terminal execution (5.0\%).

The key observation is that security-sensitive capabilities are not confined to a single server type. Web access, database access, file access, terminal execution, and content processing appear across heterogeneous application domains. Therefore, taint-style risks are not limited to obvious developer or terminal tools; many kinds of MCP tools may expose user-input-to-sensitive-operation paths.

\subsubsection{Tool Schema Pattern}

We next inspect the tool schema fields exposed during MCP registration.

\input{table/table_emp_structural.tex}

Table~\ref{tab:tool_schema_fields} shows that the required fields are consistently present: all 1,856 tools include \texttt{name}, \texttt{description}, and \texttt{inputSchema}. In contrast, optional fields are much less common. Only 160 tools use \texttt{title}, 234 use \texttt{outputSchema}, 325 use \texttt{annotations}, and 322 use \texttt{execution}; no tool uses \texttt{icons}. This suggests that MCP metadata is structurally complete at the minimum required level, but richer semantic fields are sparsely used.

We further analyze the textual content of \texttt{description} and \texttt{inputSchema/description}. We classify descriptions into four levels:

\begin{itemize}[leftmargin=*]
  \item \textit{Empty Description (T0)}: the description is empty or semantically uninformative.
  \item \textit{Basic Description (T1)}: the description only states the basic function of the tool or parameter.
  \item \textit{Constraint-Included Description (T2)}: the description includes operational constraints, such as input requirements or usage scope.
  \item \textit{Security-Aware Description (T3)}: the description explicitly includes security restrictions, misuse-prevention guidance, or access-control boundaries.
\end{itemize}

\input{table/table_emp_desc.tex}

Table~\ref{tab:mcp_tool_avg_complexity} shows that most tool descriptions remain lightweight. For the top-level \texttt{description} field, 65.19\% of tools provide only basic descriptions (T1), while only 7.00\% include security-aware information (T3). Parameter descriptions contain more operational constraints: 77.21\% of \texttt{inputSchema/description} entries are T2. However, only 1.83\% are security-aware (T3), and 9.70\% are empty.

\summary{\textbf{\textit{RQ1 Summary}:} MCP tools are diverse, and security-sensitive capabilities appear across many application domains. At the same time, existing metadata rarely provides explicit security guidance: only 7.00\% of tool descriptions and 1.83\% of parameter descriptions are security-aware. This leaves many tool categories with potential taint-style risk but limited metadata-level protection.}

\subsection{Vulnerability and Repair Property (RQ2)}

To answer RQ2, we analyze the types, triggering phases, and repair properties of the collected vulnerabilities.

\subsubsection{Vulnerability Properties}

\input{table/table_emp_vul_type.tex}

Table~\ref{tab:vulnerability} reports the distribution of vulnerability types. Command injection is the dominant category, with 27 cases (50.94\%). It is followed by path traversal with 9 cases (16.98\%), unauthorized access with 6 cases (11.32\%), DNS rebinding with 4 cases (7.55\%), SSRF with 3 cases (5.66\%), code injection with 2 cases (3.77\%), and SQL injection with 2 cases (3.77\%).

These results show that taint-style vulnerabilities dominate MCP server vulnerabilities. The typical taint-style categories in the table, including command injection, path traversal, SSRF, code injection, and SQL injection, account for 43 out of 53 cases (81.13\%). This indicates that many vulnerabilities arise when user-controlled inputs are routed into sensitive operations such as command execution, filesystem access, network requests, code execution, or database queries.

Table~\ref{tab:vulnerability} also shows that most vulnerabilities are triggered during tool invocation. Specifically, 40 cases (75.47\%) occur in the tool invocation phase, while 13 cases (24.53\%) occur during tool registration. This result highlights the central role of LLM-generated tool arguments in MCP security: unsafe behavior is often exposed when a model selects a tool and fills its parameters.

\subsubsection{Repair Properties}

We further inspect repair commits and mitigation strategies for the vulnerabilities with available remediation artifacts. We measure repair effort by the number of modified lines of code, functions, and files. We also categorize mitigation strategies following prior work on agent security~\cite{shensecurity}.

\input{table/table_emp_vul_sta.tex}

Table~\ref{tab:dev-repair-stats} shows that fixing MCP vulnerabilities often requires non-trivial code changes. On average, a fix modifies 203.6 lines of code, 5.5 functions, and 3.3 files. This suggests that remediation is not usually a local one-line patch; developers often need to understand and modify multiple parts of the server implementation.

The table also shows that most fixes rely on implementation-level changes. Secure implementation accounts for 19 cases (46.3\%), and sanitization accounts for 17 cases (41.5\%). Less common strategies include feature removal (7.3\%) and isolated environment deployment (4.9\%). Although most fixed cases are no longer exploitable (90.2\%), 4 cases (9.8\%) remain exploitable after mitigation, showing that code-level fixes can still be incomplete.

\summary{\textbf{\textit{RQ2 Summary}:} Taint-style vulnerabilities are the dominant vulnerability class in MCP servers, accounting for 81.13\% of the collected cases. Most vulnerabilities are triggered during tool invocation (75.47\%). Repairs are also complex, requiring an average of 203.6 modified lines, 5.5 functions, and 3.3 files, and 9.8\% of repaired cases remain exploitable.}

\subsection{Response to Vulnerabilities (RQ3)}

To answer RQ3, we analyze the vulnerability lifecycle and maintainer response. For each vulnerability, we collect the associated public artifacts, including issues, advisories, and fixing commits. We define the disclosure time as the earliest public report time. For fixed vulnerabilities, the fixing cycle is the time between disclosure and the fixing commit. For unpatched vulnerabilities, the exposure time is measured from disclosure to the end of our data collection period. We classify developer reactions into four categories following prior work~\cite{shensecurity}: no response, responsive but unmitigated, low-priority mitigation, and high-priority mitigation.

\input{table/table_emp_commu.tex}

Table~\ref{tab:dev-reaction-stats} shows that MCP vulnerability response is uneven and often slow. Fixed vulnerabilities take an average of 37.3 days to repair. Unpatched vulnerabilities remain exposed for an average of 92.3 days, indicating that some reported vulnerabilities persist for months.

Maintainer reaction is also mixed. High-priority mitigation accounts for 33 cases (62.3\%), suggesting that many maintainers do respond seriously to security reports. However, the remaining cases still indicate delayed or insufficient response: 9 cases (17.0\%) are responsive but unmitigated, 8 cases (15.1\%) receive low-priority mitigation, and 3 cases (5.7\%) receive no response.

\summary{\textbf{\textit{RQ3 Summary}:} MCP vulnerability remediation is slow and inconsistent. Even fixed vulnerabilities require an average fixing cycle of 37.3 days, while unpatched vulnerabilities remain exposed for 92.3 days on average. Although 62.3\% of cases receive high-priority mitigation, a substantial fraction receive delayed, low-priority, or no effective response.}

%% file: table/table_emp_structural.tex
\begin{table}[t]
\centering
\small
\setlength{\tabcolsep}{6pt}
\begin{tabular}{lrrr}
\toprule
\textbf{Field} & \textbf{Used Count} & \textbf{Empty Count} \\
\midrule
\texttt{name}        & 1856 & 0     \\
\texttt{title}       & 160  & 1696  \\
\texttt{description} & 1856 & 0     \\
\texttt{icons}       & 0    & 1856  \\
\texttt{inputSchema} & 1856 & 0     \\
\texttt{outputSchema}& 234  & 1622  \\
\texttt{annotations} & 325  & 1531  \\
\texttt{execution}   & 322  & 1534  \\
\bottomrule
\end{tabular}
\caption{Usage statistics of tool schema fields in 1,856 MCP tools.}
\label{tab:tool_schema_fields}
\end{table}

%% file: table/table_emp_desc.tex
\begin{table}[!t]
\centering
\scriptsize
\setlength{\tabcolsep}{3pt}
\resizebox{\columnwidth}{!}{%
\begin{tabular}{m{1.5cm}m{0.7cm}m{0.7cm}m{0.7cm}m{0.7cm}m{0.7cm}m{0.7cm}} 
\toprule
\textbf{Field} & \textbf{Level} & \textbf{MCP Tool (\#)} & \textbf{Per. (\%)} & \textbf{Tokens (\#)} & \textbf{Sentence No. (\#)} & \textbf{Avg. Sent. Len.} \\
\midrule
\multirow{5}{*}{\textit{Description}} 
& T0 & 1    & 0.05  & 0.00   & 0.00  & 0.00  \\
& T1 & 1210 & 65.19 & 14.56  & 1.26  & 11.12 \\
& T2 & 515  & 27.75 & 97.33  & 4.61  & 23.02 \\
& T3 & 130  & 7.00  & 263.78 & 9.39  & 32.67 \\
\cmidrule(lr){2-7}
& Total & 1856 & 100.00 & 54.97 & 2.76 & 15.92 \\
\hline
\multirow{5}{*}{\shortstack[l]{\textit{inputSchema/}\\\textit{Description}}}
& T0 & 180  & 9.70  & 0.00   & 0.00  & 0.00  \\
& T1 & 209  & 11.26 & 25.63  & 1.45  & 19.64 \\
& T2 & 1433 & 77.21 & 117.79 & 5.77  & 25.01 \\
& T3 & 34   & 1.83  & 315.21 & 13.53 & 29.97 \\
\cmidrule(lr){2-7}
& Total & 1856 & 100.00 & 99.61 & 4.86 & 22.07 \\
\bottomrule
\end{tabular}%
}
\caption{MCP Tool Textual Patterns}
\label{tab:mcp_tool_avg_complexity}
\end{table}

%% file: table/table_emp_vul_type.tex
\begin{table}[!t]
\centering
\small
\resizebox{\columnwidth}{!}{%
\begin{tabular}{m{2cm}m{3cm}m{1cm}m{1cm}} 
\toprule
\textbf{Dimension} & \textbf{Category} & \textbf{No.} & \textbf{Per. (\%)}\\ 
\midrule
\multirow{6}{*}{\begin{tabular}[c]{c@{}c@{}}Vulnerability \\ Type\end{tabular}} 
                  & Command Injection & 27 & 50.94 \\
                  & Path Traversal & 9 & 16.98 \\ 
                  & Unauthorized Access & 6 & 11.32 \\
                  & DNS Rebinding & 4 & 7.55 \\
                  & SSRF & 3 & 5.66 \\ 
                  & Code Injection & 2 & 3.77 \\
                  & SQL Injection & 2 & 3.77 \\
                  \hline
\multirow{2}{*}{\begin{tabular}[c]{c@{}c@{}}Triggering\\Phase\end{tabular}} 
                  & Tool Invocation & 40 & 75.47 \\
                  & Tool Registration & 13 & 24.53 \\
                  \bottomrule
\end{tabular}%
}
\caption{Overall Statistics of MCP and Vulnerability Type}
\label{tab:vulnerability}
\end{table}

%% file: table/table_emp_vul_sta.tex
\begin{table}[!t]
\centering
\small
\setlength{\tabcolsep}{6pt}
\resizebox{\columnwidth}{!}{%
\begin{tabular}{m{1.5cm}m{3.5cm}m{1cm}m{0.7cm}} 
\toprule
{\textbf{Dimension}} & \textbf{Category} & \textbf{No.} & \textbf{Per. (\%)} \\ 
\midrule
\multirow{3}{*}{\shortstack[l]{Repair \\ Effort}}
& Average Lines of Code & 203.6 &-- \\
& Average Modified Functions & 5.5 &--\\
& Average Modified Files & 3.3 &--\\ 

\midrule
\multirow{4}{*}{\shortstack[l]{Mitigation \\ Strategy}} 
                  & Secure Implementation   & 19  & 46.3 \\
                  & Sanitization             & 17  & 41.5 \\
                  & Feature Removal          & 3   & 7.3  \\
                  & Isolated Environment     & 2   & 4.9  \\ 
                  \midrule
\multirow{2}{*}{\shortstack[l]{Mitigation \\ Effectiveness}} 
                  & No Longer Exploitable    & 37  & 90.2 \\
                  & Still Exploitable        & 4   & 9.8  \\ 
                  \bottomrule
\end{tabular}%
}
\caption{Repair Statistics of Vulnerable MCP Tools}\label{tab:dev-repair-stats}
\end{table}

%% file: table/table_emp_commu.tex
\begin{table}[!t]
\centering
\small
\setlength{\tabcolsep}{6pt}
\resizebox{\columnwidth}{!}{%
\begin{tabular}{m{1.5cm}m{4cm}m{1cm}m{0.7cm}} 
\toprule
\textbf{Dimension} & \textbf{Metric/Category} & \textbf{Value} & \textbf{Per.} \\ \midrule
\multirow{2}{*}{\shortstack[l]{Vulnerability \\ Lifespan}} 
                  & Fixing Cycle      & 37.3d  & -- \\
                  & Exposure Time     & 92.3d & -- \\ \midrule

\multirow{4}{*}{\shortstack[l]{Developer \\ Reaction}}
                  & High-Priority Mitigation       & 33 & 62.3 \\
                  & Responsive but Unmitigated      & 9  & 17.0 \\
                  & Low-Priority Mitigation         & 8  & 15.1 \\
                  & No Response                     & 3  & 5.7  \\
                   \bottomrule
\end{tabular}%
}
\caption{Statistics of Community Reaction to Vulnerable MCP Tools}
\label{tab:dev-reaction-stats}
\end{table}

%% file: src/ch04-approach.tex
\section{\tool}\label{sec:approach}

We present \tool, a lightweight defense for mitigating taint-style vulnerability exploitation in MCP tools without modifying MCP Server implementations. \tool operates at the MCP interaction layer: it estimates potential risks from tool metadata, augments tool descriptions with security-aware constraints, and introduces reflection around tool invocation.


\subsection{Motivation and Design Overview}

MCP tools may expose sensitive capabilities such as file access, command execution, code execution, database queries, network requests, data parsing, identity and access control, and environment-variable access. When user-controlled inputs are passed to these capabilities, the tool may be abused through taint-style vulnerabilities such as OS command injection, code injection, SQL injection, path traversal, and SSRF.

\tool is designed for cases where the MCP Server code is unavailable, difficult to audit, or undesirable to modify. Instead of patching the server implementation, \tool strengthens the interface between the LLM and MCP tools through three lightweight steps. First, it estimates which tools and parameters may involve taint-style risks. Second, it rewrites tool descriptions with explicit security constraints in the offline stage. Third, it asks the LLM to reflect on planned or completed tool calls in the online stage before unsafe behavior is propagated.

The goal of \tool is not to prove that a tool is vulnerable. Rather, it provides a conservative warning signal and uses this signal to guide LLM tool-use behavior. This design treats MCP metadata as a lightweight policy surface that can influence tool selection, parameter generation, and refusal decisions.

\subsection{Risk Identification and Taint-style Vulnerability Estimation}

Given the metadata of an MCP tool, denoted as $\mathbf{M}$, \tool estimates a coarse-grained risk profile:
\[
    \mathbf{R} = \langle \mathcal{C}, \mathcal{P}, \mathcal{W} \rangle,
\]
where $\mathcal{C}$ denotes high-risk capabilities, $\mathcal{P}$ denotes potentially tainted parameters, and $\mathcal{W}$ denotes candidate CWE categories.

This step uses the tool name, description, and input schema to answer three simple questions. First, does the tool expose a high-risk capability, such as filesystem access, command or code execution, database access, network access, parsing, credential access, or authorization control? Second, which parameters may carry user-controlled input, such as paths, commands, code snippets, queries, URLs, headers, templates, filters, or identifiers? Third, which taint-style CWE risks may arise when those parameters affect the corresponding capability?

The output is intentionally conservative. \tool does not claim that the implementation is exploitable; it only estimates whether the metadata suggests a plausible user-input-to-sensitive-operation pattern.

\subsection{Tool Metadata Augmentation}

In the offline stage, \tool rewrites the \texttt{description} field of each risky tool while preserving the original tool interface and server code. The augmented description keeps the functional meaning of the tool, but adds security information that the LLM should consider during planning.

The augmented description contains four concise elements:
\begin{itemize}[leftmargin=*]
    \item \textbf{Risk capability.} What sensitive operation the tool can perform.
    \item \textbf{Tainted parameters.} Which arguments may contain user-controlled input.
    \item \textbf{Potential CWE risks.} Which taint-style vulnerability classes may be relevant.
    \item \textbf{Invocation policy.} Which intents or input patterns should be rejected, and what checks should be performed before calling the tool.
\end{itemize}

This makes the security boundary explicit in natural language. Since LLMs rely on tool descriptions when selecting tools and filling parameters, embedding these constraints in metadata can discourage calls that route untrusted input into sensitive operations.

\subsection{Tool Invocation Reflection}

In the online stage, \tool adds a reflection step around tool invocation. Before executing a tool call, the LLM is asked to reconsider the original user intent, the selected tool, and the proposed arguments. The reflection checks whether the call is within the tool's intended scope, whether the user appears authorized, and whether the arguments may trigger the risks described in the augmented metadata.

If the reflection identifies a potential exploit attempt or an unsafe parameter, the invocation is refused or revised before it reaches the MCP Server. \tool may also apply reflection after a call when the returned result could influence later tool use, preventing unsafe multi-step escalation.

Together, metadata augmentation and invocation reflection provide a defense that is simple to deploy, does not require server-side code changes, and directly targets the LLM-mediated tool-use process in MCP.

%% file: src/ch05-dataset.tex
\section{Benchmark of Malicious MCP Attacks}\label{sec:dataset}

Based on the 53 vulnerabilities collected in our empirical study, we construct a benchmark of malicious MCP attacks to evaluate whether LLM agents can be induced to exploit taint-style vulnerabilities exposed by MCP servers. Each case is grounded in a real vulnerable tool and preserves the corresponding user-controlled parameter, sensitive operation, and attack goal. The benchmark therefore measures exploit-triggering behavior at the MCP tool interface, rather than general policy violations detached from server-side vulnerabilities.

For each vulnerability, we design trigger prompts that exercise the corresponding vulnerable data flow and define attack objectives inspired by MITRE ATT\&CK tactics~\cite{mitre_attack}. To improve prompt diversity and adversarial realism, we augment the benchmark with jailbreak strategies adapted from Liu et al.~\cite{liu2023jailbreaking}; these variants change the surface form of the malicious intent while preserving the same tool-level risk. In total, our benchmark contains 792 attack prompts, covering five representative taint-style attack objectives, ten jailbreak strategies, five categories of taint-style vulnerabilities, 45 MCP servers, and 130 MCP tools.

%% file: src/ch06-evaluation.tex
\section{Experiment}\label{sec:evaluation}

\input{table/table_eval_effective}

We evaluate \tool with the following research questions.
\begin{itemize}[leftmargin=*]
\item \textbf{RQ1: Effectiveness.} How effective is \tool in mitigating taint-style vulnerability exploitation in MCP servers?
\item \textbf{RQ2: Ablation Study.} How does each component of \tool contribute to the mitigation effect?
\end{itemize}

\subsection{Setup}

\textbf{Dataset.}
We use the benchmark described in Section~\ref{sec:dataset}. It contains 792 malicious MCP attack prompts, covering five taint-style attack objectives, five categories of taint-style vulnerabilities, 45 MCP servers, and 130 MCP tools.

\textbf{LLM.}
We use GPT-4o~\cite{openai2024gpt4o} as the evaluation model. GPT-4o has strong instruction-following and tool-use capabilities, making it suitable for evaluating whether security-aware metadata and reflection can guide MCP tool invocation.

\textbf{Settings.}
Table~\ref{tab:final-asr-results} compares four metadata settings. \textit{None} denotes no prompt-level defense. \textit{Decl.} denotes a generic declaration of taint-style vulnerability risks. \textit{Wrong} denotes an incorrect taint-style vulnerability description. \textit{Ident.} denotes the vulnerability risk identified by \tool and embedded into the tool description. We further compare two invocation settings: \textit{Pre-reflection}, which evaluates the model before applying tool invocation reflection, and \textit{Post-reflection}, which evaluates the model after applying reflection.

\textbf{Metrics.}
We report two attack success rates. Trial-level attack success rate ($ASR_{trial}$) measures the fraction of successful attack attempts:
\[
ASR_{trial} = \frac{\#\text{Successful Attack Trials}}{\#\text{Total Attack Trials}}.
\]
Case-level attack success rate ($ASR_{case}$) measures the fraction of cases in which at least one attack succeeds:
\[
ASR_{case} = \frac{\#\text{Cases with At Least One Successful Attack}}{\#\text{Total Test Cases}}.
\]
For both metrics, lower values indicate stronger mitigation.

\subsection{RQ1 Effectiveness}

Table~\ref{tab:final-asr-results} shows that MCP agents are highly vulnerable without defense. Under the \textit{None} and \textit{Pre-reflection} setting, the attack success rate reaches 56.61\% at the trial level and 63.89\% at the case level.

\tool substantially reduces attack success. With identified risk descriptions and invocation reflection, the attack success rate drops to 0.04\% at the trial level and 0.13\% at the case level. Compared with the undefended setting, this corresponds to reductions of 56.57 and 63.76 percentage points, respectively. The near-zero case-level ASR indicates that \tool remains effective even when attacks are evaluated at the case level.

\noindent\textbf{Answer to RQ1.}
\tool effectively mitigates taint-style vulnerability exploitation in MCP servers, reducing the final attack success rate to 0.04\% at the trial level and 0.13\% at the case level.

\subsection{RQ2 Ablation Study}

We next examine the contribution of each component. First, metadata augmentation is useful even without reflection. In the \textit{Pre-reflection} row, a generic declaration lowers $ASR_{trial}$ from 56.61\% to 21.63\%, and lowers $ASR_{case}$ from 63.89\% to 28.41\%.

Second, risk identification improves the quality of the augmented description. Under \textit{Pre-reflection}, the identified risk description further reduces $ASR_{trial}$ to 6.69\% and $ASR_{case}$ to 9.09\%, outperforming both the generic declaration and the incorrect vulnerability description. Under \textit{Post-reflection}, the identified risk description also achieves the lowest ASR among all metadata settings.

Third, invocation reflection provides an additional reduction across all metadata settings. Even without metadata augmentation, reflection lowers $ASR_{trial}$ from 56.61\% to 2.19\% and $ASR_{case}$ from 63.89\% to 3.16\%. When combined with the identified risk description, reflection further reduces the rates from 6.69\% to 0.04\% and from 9.09\% to 0.13\%.

The best result is achieved by combining all components. Under \textit{Post-reflection}, the identified risk description outperforms both the generic declaration (0.67\% trial-level ASR and 1.52\% case-level ASR) and the incorrect vulnerability description (0.17\% trial-level ASR and 0.51\% case-level ASR). This suggests that reflection is powerful, but risk-specific metadata still provides useful guidance.

\noindent\textbf{Answer to RQ2.}
All three components contribute to the final mitigation effect. Metadata augmentation reduces unsafe tool planning, risk identification makes the guidance more precise, and invocation reflection further filters unsafe calls.

%% file: table/table_eval_effective.tex
\begin{table*}[!t]
\centering
\small
\setlength{\tabcolsep}{5pt}
\begin{tabular}{lcccccccc}
\toprule
\multirow{2}{*}{\textbf{Setting}} 
& \multicolumn{2}{c}{\textbf{None}} 
& \multicolumn{2}{c}{\textbf{Decl.}} 
& \multicolumn{2}{c}{\textbf{Wrong}} 
& \multicolumn{2}{c}{\textbf{Ident.}} \\
\cmidrule(lr){2-3} \cmidrule(lr){4-5} \cmidrule(lr){6-7} \cmidrule(lr){8-9}
& \multicolumn{1}{c}{$\mathbf{ASR}_{\mathbf{trial}}$} 
& \multicolumn{1}{c}{$\mathbf{ASR}_{\mathbf{case}}$}
& \multicolumn{1}{c}{$\mathbf{ASR}_{\mathbf{trial}}$} 
& \multicolumn{1}{c}{$\mathbf{ASR}_{\mathbf{case}}$}
& \multicolumn{1}{c}{$\mathbf{ASR}_{\mathbf{trial}}$} 
& \multicolumn{1}{c}{$\mathbf{ASR}_{\mathbf{case}}$}
& \multicolumn{1}{c}{$\mathbf{ASR}_{\mathbf{trial}}$} 
& \multicolumn{1}{c}{$\mathbf{ASR}_{\mathbf{case}}$} \\
\midrule
Pre-reflection  & 56.61\% & 63.89\% & 21.63\% & 28.41\% & 9.89\% & 12.75\% & 6.69\% & 9.09\% \\
Post-reflection & 2.19\%  & 3.16\%  & 0.67\%  & 1.52\%  & 0.17\% & 0.51\%  & 0.04\% & 0.13\% \\
\bottomrule
\end{tabular}
\caption{Final attack success rates under different prompt defense settings. 
None denotes no prompt defense; Decl. denotes a generic declaration of taint-style vulnerabilities; 
Wrong denotes an incorrect taint-style vulnerability description; 
Ident. denotes the taint-style vulnerability identified by our method.}
\label{tab:final-asr-results}
\end{table*}

%% file: src/ch07-related-work.tex

\section{Related Work}

\subsection{MCP Ecosystem and Security}

Recent work has begun to study MCP as an emerging infrastructure for tool-augmented LLM agents. MCPCorpus~\cite{lin2025large} and MCPZoo~\cite{wu2025mcpzoolargescaledatasetrunnable} collect large-scale MCP servers and clients, enabling ecosystem-level measurement and runtime evaluation. Several benchmarks further evaluate MCP-based agents from different perspectives, including tool-use capability~\cite{song2025help}, multi-step workflows~\cite{wu2025mcpmark}, security robustness~\cite{yang2025mcpsecbench,zhang2025mcp}, and multi-tool interaction~\cite{fan2025mcptoolbench++}. These studies provide important infrastructure for MCP evaluation, while our work focuses on real-world disclosed vulnerabilities and their mitigation. In particular, our benchmark is not only a collection of adversarial prompts; each attack is tied to a concrete vulnerable MCP tool and an observed taint-style risk.

Security studies of MCP mainly examine malicious or unsafe MCP interactions. Prior work identifies attacks such as tool poisoning, puppet attacks, rug-pull attacks, malicious external resources, and preference manipulation~\cite{song2025beyond,wang2025mpma}. Other studies measure structural risks in the MCP ecosystem~\cite{guo2025measurement}, analyze description-code inconsistency~\cite{li2026don}, or use static and MCP-specific analysis to uncover server-side security issues~\cite{hasan2025model}. Complementary to these efforts, we collect real vulnerability reports, analyze their metadata, triggering phases, repair properties, and community responses, and show that taint-style vulnerabilities are a dominant risk in MCP servers.

Several defense mechanisms have also been proposed for MCP-based agents. Jing et al.~\cite{jing2025mcip} introduce a contextual-integrity protocol for safer LLM-MCP interactions. AgentBound~\cite{buhler2025securing} and MCP Guardian~\cite{kumar2025mcp} focus on access control, authentication, monitoring, and system-level safeguards. MCP-Guard~\cite{xing2025mcp} combines static scanning, neural detection, and LLM-based arbitration. Our work differs in threat model and intervention point: we assume MCP servers may be benign but vulnerable, and mitigate exploit-triggering tool calls by augmenting tool metadata and adding invocation reflection, without modifying server implementations.

This distinction is important because vulnerable MCP servers expose a different risk surface from malicious MCP servers. In the malicious-server setting, the primary concern is whether a server can poison the LLM's context, manipulate tool selection, or return adversarial content. In our setting, the server may be honestly implemented for its intended task, but still mishandle user-controlled parameters. Therefore, the defense must guide the LLM before it sends risky arguments to the tool. \tool addresses this gap by treating metadata as a security-relevant interface, not merely as functional documentation.

\subsection{Taint-Style Vulnerabilities in LLM Agents}

Taint-style vulnerabilities occur when untrusted input flows into sensitive operations. Traditional research studies such vulnerabilities across C programs, OS kernels, web applications, scripting languages, firmware, and microservice systems~\cite{yamaguchi2015automatic,zhang2021statically,luo2022tchecker,kang2023scaling,gibbs2024operation,liu2025detecting}. These works mainly focus on program analysis, vulnerability detection, or exploit validation in conventional software.

Recent work extends this problem to LLM-integrated systems. Shen et al.~\cite{shensecurity} show that LLM agents can expose code injection, SQL injection, command injection, path traversal, and related vulnerabilities. LLMSmith~\cite{liu2024demystifying}, prompt-to-SQL injection studies~\cite{pedro2025prompt}, and AgentFuzz~\cite{liu2025make} further demonstrate that LLM-generated inputs can trigger vulnerable data flows in agent frameworks and applications. These studies show that LLMs may become a new source of untrusted inputs for existing software sinks. Our work focuses specifically on MCP servers and studies how the MCP metadata and tool invocation process can be used as a complementary defense point for reducing such exploit-triggering behavior.

\subsection{LLM Jailbreaking and Defenses}

LLM jailbreaking aims to bypass model safety constraints and induce unsafe outputs. Existing attacks include white-box optimization such as GCG~\cite{zou2023universal,jia2024improved}, black-box optimization such as PAIR, TAP, GPTFuzz, AutoDAN, and PaPillon~\cite{chao2025jailbreaking,mehrotra2024tree,yu2023gptfuzzer,liu2023autodan,gong2025papillon}, as well as social-engineering and encoding-based strategies~\cite{johnson2024generation,yang2024dark,zeng2024johnny,jiang2024artprompt,yuan2023gpt,yong2023low}. Empirical studies further characterize jailbreak prompt patterns and their evolution~\cite{liu2023jailbreaking,yu2024don,shen2024anything}. We draw on these taxonomies to construct diverse malicious MCP prompts, but our target is not general policy violation; it is the exploitation of vulnerable MCP tools through LLM-mediated tool invocation.

Defenses against jailbreaking commonly rely on input preprocessing, output filtering, and prompt engineering. Representative approaches include retokenization and smoothing~\cite{jain2023baseline,robey2023smoothllm}, certified erase-and-check defenses~\cite{kumar2023certifying}, intent analysis~\cite{zhang2025intention}, safety classifiers such as Llama Guard and Prompt Guard~\cite{meta_llama_guard4,meta_llama_prompt_guard}, self-examination or self-refinement~\cite{phute2308self,kim2025break}, multi-agent filtering~\cite{zeng2024autodefense}, and safety-oriented prompting~\cite{wu2023defending,zhang2024defending,wei2023jailbreak}. \tool is related to prompt-based and reflection-based defenses, but applies them at the MCP tool interface: security guidance is embedded into tool descriptions, and reflection is used to reassess concrete tool invocations before they reach vulnerable server-side code.

Compared with general jailbreak defenses, MCP tool use introduces structured arguments and explicit tool metadata. A prompt may appear benign at the natural-language level while still causing unsafe behavior through a path, URL, command, query, or code parameter. This makes MCP security different from ordinary refusal-based safety filtering. Our defense therefore focuses on the relation between user intent, tool capability, and parameter semantics.

%% file: src/ch08-conclusion.tex

\section{Conclusion}

This paper presents a systematic study of MCP server vulnerabilities and shows that taint-style vulnerabilities are a major security risk in the MCP ecosystem. Our empirical results reveal that security-sensitive capabilities appear across diverse tool categories, while existing tool metadata rarely provides explicit security guidance. We further find that taint-style vulnerabilities account for a dominant fraction of collected vulnerabilities, are mostly triggered during tool invocation, and often require non-trivial code changes and long response cycles to repair.

Motivated by these findings, we propose \tool, a lightweight defense that mitigates taint-style vulnerability exploitation without modifying MCP Server implementations. \tool estimates potential taint-style risks from tool metadata, augments tool descriptions with security-aware constraints, and applies invocation reflection to reconsider unsafe tool calls. Our evaluation shows that \tool reduces the final attack success rate to 0.04\% at the trial level and 0.13\% at the case level. These results suggest that security-aware metadata and LLM reflection provide a practical, deployable, and complementary direction for securing MCP-based agents.